\documentclass[twocolumn,prd,superscriptaddress,longbibliography]{revtex4-2}
\pdfoutput=1
\usepackage{bm}
\usepackage{graphicx}
\usepackage{booktabs}
\usepackage{slashed}
\usepackage{hyperref}
\usepackage{amssymb,amsmath,amsfonts,braket}
\usepackage{MnSymbol}
\usepackage{color}
\usepackage[utf8]{inputenc}
\usepackage[caption=false]{subfig}

\def \beq  {\begin{equation}}
\def \eeq  {\end{equation}}
\def \beqar {\begin{eqnarray}}
\def \eeqar {\end{eqnarray}}
\def\sqr#1#2{{\vcenter{\vbox{\hrule height.#2pt
\hbox{\vrule width.#2pt height#1pt \kern#1pt
\vrule width.#2pt}\hrule height.#2pt}}}}

\def\vp {{\vec p}}
\def\vq {{\vec q}}

\def\Tr {{\rm Tr}}

\def\vu {\vec{u}}

\def\e {\epsilon}

\def\Z{{\mathcal Z}}

\begin{document}

\title{Spectral functions at nonzero temperature}

\author{V. P. Nair}
\affiliation{Department of Physics, City College of New York, CUNY, New York, NY 10031}
\author{Robert D. Pisarski}
\affiliation{Department of Physics, Brookhaven National Laboratory, Upton, NY 11973}

\begin{abstract}
We present a straightforward derivation of the spectral representation of
a scalar field at nonzero temperature, assuming that
the field is relativistically invariant in vacuum.  This form
was first derived by Bros and Buchholz
\cite{Bros:1992ey,Bros:1994ofl,Bros:1996mw,Bros:2001zs}.
\end{abstract}

\maketitle
\section{Introduction}
\label{sec:intro}
The analysis of field theories is of interest in a variety of physical
problems, from the behavior of the early universe, to the collisions of
heavy ions.  A fundamental quantity of interest is the behavior of spectral
functions, from which transport quantities can be computed using the Kubo
formula.  

In vacuum, by relativistic invariance
the spectral function is a function of a single variable, which can be chosen
to be the Mandelstam variable $s$.  At nonzero temperature, because relativistic
invariance is lost, generally one expects that the spectral density is then
a function of two variables, such as $s$ and the spatial momentum, $p^2$.  

For a general theory at nonzero temperature, one does not expect any specific
relation in the spectral density between $s$ and $p^2$.  By developing a set of axioms, similar to the Wightman axioms, but for fields at nonzero temperature,
Bros and Buchholz showed that 
{\it if} the theory is relativistically invariant in vacuum, then
that constrains the form of the spectral density at nonzero temperature.
The key point is that the difference between the vacuum correlations
and those at nonzero temperature is in the choice of the state.
If the underlying dynamics is relativistically invariant,
then this informs the spectral representation even at nonzero
temperature.

In this paper we present an alternate derivation of the Bros-Buchholz form in a more
straightforward manner, without the use of axiomatic field theory.  
The utility of the Bros-Buchholz form has been emphasized
recently by Lowdon {\it et al.} \cite{Lowdon:2021ehf,Lowdon:2022keu,Lowdon:2022xcl,Lowdon:2022ird,Lowdon:2025fyb,Bala:2023iqu}.

\section{The Bros-Buchholz Representation}
\label{sec:BB}

We construct the Bros-Buchholz representation at nonzero temperature along the lines of how the standard K\"allen-Lehmann representation is obtained in field theory.
We consider a scalar field theory and start with the definition of the Wightman function at nonzero temperature,
\beq
W(x, y) = {1\over \Z } \Tr \left[ e^{- \beta H} \phi (x) \phi (y)\right] \; .
\label{BB1}
\eeq
As usual, using the spacetime translational invariance at the level of operators
we can write
\beq
\phi(x) = e^{ i P\cdot x} \, \phi(0) \, e^{- i P\cdot x} \; ,
\label{BB2}
\eeq
where $P_\mu$ is the total 4-momentum operator. Using this result and
a complete set of states $\ket{a}$ and $\ket{b}$,
\beq
W(x,y) = {1\over \Z} \sumint_{a,b} e^{-\beta E_b} \vert \phi_{ab}(0)\vert^2
e^{- i (p_a-p_b)\cdot (x-y) } \; .
\label{BB3}
\eeq
The states $\ket{a}$ and $\ket{b}$ are characterized by their energy 
$\epsilon_\alpha$ in its rest frame and the 3-momentum $\vp_a$ so that
$\phi_{ab}(0) = \bra{\alpha, \vp_a} \phi(0) \ket{\beta, \vp_b}$.
The sum-integral in Eq. (\ref{BB3}) represents
\beqar
\sumint_a &=& \sum_\alpha \int {d^3 p_a \over 2 p_a^0(2\pi)^3} ,
\hskip .2in p_a^0 = \sqrt{\e_\alpha^2 + \vp_a^2}\nonumber\\ 
&=& \sum_\alpha \int_0^\infty d \sqrt{s'} \,\delta (\sqrt{s'} - \e_\alpha )
\int 
{d^3 p_a \over 2 p_a^0 (2\pi)^3} \; ,
\label{BB4}
\eeqar
where, in the last line, $p_a^0 = \sqrt{s' + \vp_a^2}$.
We can thus write Eq. (\ref{BB3}) as
\beqar
W(x,y)&=& {1\over \Z} \int \frac{d^4 p}{(2 \pi)^4}\, e^{- i p\cdot (x-y)} 
\sum_{\alpha,\beta} \int_0^\infty {ds'\over 2 \sqrt{s'}} 
\delta (\sqrt{s'} - \e_\alpha )
 \label{BB5}\nonumber \\
&&
\int_0^\infty {ds''\over 2 \sqrt{s''}}  \delta (\sqrt{s''} - \e_\beta )
\int {d^3 p_a \over 2 p_a^0 (2\pi)^3}\\
&&~~
\int 
{d^3 p_b \over 2 p_b^0(2\pi)^3}
e^{-\beta p_b^0}  \vert \phi_{ab}(0)\vert^2 \;\delta^{(4)} (p - p_a + p_b) , \nonumber
\eeqar
where $p_b^0 = \sqrt{s'' +\vp_b^2}$.

We now note that, since the Fourier transform of $\phi(0)$ is an integral
over all momenta, the matrix element $\phi_{ab}(0)$ in Eq. (\ref{BB5})
can be nonzero for any values of $p_a- p_b$, and so is a function of
$\vp_a - \vp_b$, $\e_\alpha$, and $\e_\beta$.
Further, from the Lorentz transformation property of $\phi$, we have
\beq
\phi (\Lambda^{-1} x) = U(\Lambda)\, \phi (x) \, U^{-1} (\Lambda) \; .
\label{BB6}
\eeq
Setting $x = 0$, we see that
\beq
\bra{a} \phi (0) \ket{b} = \bra{a} U(\Lambda) \, \phi(0)\, U^{-1} (\Lambda) \ket{b}
= \bra{a'} \phi (0) \ket{b'} \; ,
\label{BB7}
\eeq
where $\ket{a'}$, $\ket{b'}$ are the Lorentz transforms of
$\ket{a}$, $\ket{b}$. We see that the value of the matrix element
is unaltered under the Lorentz transformation of
$\ket{a}$, $\ket{b}$.
We can therefore take the matrix element to be a function
of the invariants $(p_a- p_b)^2$ and $\e_\alpha$, $\e_\beta$.
So we can define an invariant function
\beqar
&&f(s', s'', (p_a - p_b)^2)=\nonumber\\
&&\hskip .2in \sum_{\alpha, \beta} \delta(\sqrt{s'} - \e_\alpha) 
\delta(\sqrt{s''} -\e_\beta) { \vert \bra{a} \phi(0)\ket{b}\vert^2 \over 4 \sqrt{s'} \sqrt{s''}}
\label{BB8}
\eeqar
Using this relation, we can now write Eq. (\ref{BB5}) as
\beqar
W(x, y) &=& {1\over \Z} 
\int \frac{d^4 p}{(2 \pi)^4} \,e^{-i p\cdot (x-y)}  
\int {d^3p_a \over 2 p_a^0 (2\pi)^3}
\nonumber\\
&&\int {d^3p_b \over 2 p_b^0 (2\pi)^3}
\delta^{(4)}(p - p_a + p_b)
\label{BB9}\\
&& \int_0^\infty
ds' 
\int_0^\infty
ds''\,e^{-\beta p_b^0} f( s', s'', (p_a - p_b)^2) \; . \nonumber
\eeqar
The last argument of $f$ can be written out as
\beqar
(p_a - p_b)^2 &=& s' + s'' - 2 p_a \cdot p_b\label{BB10}\\
&=& s' + s'' - 2 \sqrt{\vp_a^2 + s'} \, \sqrt{\vp_b^2+ s''} \, 
+ 2 \vp_a \cdot\vp_b \; .
\nonumber
\eeqar

Consider now the integral
\beqar
{\mathcal I} &=& \int_0^\infty ds' \int {d^3p_a \over 2 p_a^0 (2\pi)^3} 
\,\delta^{(4)}(p - p_a + p_b)
\nonumber\\
&&\times
f( s', s'', (p_a - p_b)^2) 
\\
&=&\int_0^\infty ds' \, \int {d^3p_a \over 2 p_a^0 (2\pi)^3}
 \delta (p^0 - p_a^0 + p_b^0)
\delta^{(3)}(\vq - \vp_a)\label{BB11}\nonumber\\
&&\times 
f( s', s'', s' + s'' - 2 \sqrt{\vp_a^2 + s'} \, \sqrt{\vp_b^2+ s''} \, + 2 \vp_a \cdot\vp_b ) \; . \nonumber 
\eeqar
where $\vq = \vp + \vp_b$.  
We can then use the $\delta$-function to replace the integral over
$\vp_a$ by one over $\vq$, so that ${\mathcal I}$
is a function of $\vp_b$, $\vq$, $p^0$, and $s''$. 

The essential point is then that both the
arguments of $f$, and the integration measures, are Lorentz-invariant.
Thus ${\mathcal I}$ must be a function of the Lorentz invariants constructed
from the variables $\vp_b$, $\vq$, $p^0$, and $s''$. 
Since $p^0$ is the only
time-component, there are five possible invariants:
\beqar
&&\alpha_1 = (p^0)^2 - \vq^2 \; \; , \; \; 
\alpha_2 = (p^0)^2 - \vp_b^2 \; \; , \nonumber\\
&&\alpha_3 = (p^0)^2 - \vp_b\cdot \vq \; \; , \; \; 
\alpha_4 = s'' \; \; , \nonumber\\
&&\epsilon (p^0) = \begin{cases}
+1 &p^0>0\\
-1&p^0<0\\
\end{cases} \; .
\label{BB12}
\eeqar
(We could consider $\theta(p^0)$ and $\theta(-p^0)$ as well, where $\theta$ denotes the step-function, but these are equivalent to what we have in Eq. (\ref{BB12}) since
$\theta (\pm p^0) = {1\over 2} ( 1\pm \epsilon (p^0))$.
Thus we can write
\beq
{\mathcal I} = {\mathcal I}_1(\alpha_1, \alpha_2, \alpha_3, \alpha_4)
+ \epsilon(p^0)\, {\mathcal I}_2(\alpha_1, \alpha_2, \alpha_3, \alpha_4)
\; . \label{BB13}
\eeq

Now consider the special case when the state $\ket{b}$ is
the vacuum state. In this case, $\vp_b = 0$ and 
$\epsilon_\beta = 0$, so that, once the integration
over $s''$ is carried out, it is set to zero.
The corresponding ${\mathcal I}$ takes the form
\beqar
{\mathcal I} \big\vert_{\vp_b = 0, s'' = 0}
&=& {\mathcal I}_1 \bigl((p^0)^2 - \vp^2, (p^0)^2 , (p^0)^2, 0\bigr))\label{BB14}\\
&&
+ \epsilon(p^0) {\mathcal I}_2 \bigl((p^0)^2 - \vp^2, (p^0)^2 , (p^0)^2, 0\bigr))
\; .
\nonumber
\eeqar
If the state $\ket{b}$ is the vacuum state, the
spectral representation should reduce to the usual K\"allen-Lehmann
representation, which is of the form \cite{Itzykson:1980rh}
\beq
W(x,y) = \int {d^4 p \over (2\pi)^4} e^{-i p\cdot(x-y)}
\theta (p^0) \, \rho (p^2)
\label{BB14a}
\eeq
for some spectral density $\rho(p^2)$.
Thus, functions on the right-hand-side of Eq. (\ref{BB14}) can only
depend on $\epsilon(p^0)$ and $(p^0)^2 - \vp^2$.
The compatibility of
the form of ${\mathcal I}$ in Eq. (\ref{BB13})
with the form obtained for the particular
case of $\ket{b} = \ket{0}$
shows that, in general, there should be no dependence on the variables in the second and
third variables, $\alpha_2$ and $\alpha_3$, in ${\mathcal I}$.
We therefore conclude that, at {\it any} temperature, 
\beq
{\mathcal I} = {\mathcal I}_1 \bigl( (p^0)^2 - \vq^2, s''\bigr)
+ \epsilon(p^0) \, {\mathcal I}_2 \bigl( (p^0)^2 - \vq^2, s''\bigr)
\; .
\label{BB15}
\eeq
Having argued that there is no dependence on $\alpha_2$ and
$\alpha_3$, we can see that the functions
${\mathcal I}_1$ and ${\mathcal I}_2$ at finite temperature can be obtained
from the corresponding vacuum ones with the shift
$\vp \rightarrow \vq = \vp +\vp_b$.
 Physically, this just reflects that fact that in
the trace which enters into the partition function at nonzero temperature, Eq. (\ref{BB1})
and (\ref{BB3}), is a sum over all states, including those with momenta $\vp_b$, and not just the vacuum.

The second point is that the K\"allen-Lehmann representation also shows that
we need 
${\mathcal I}_1 ((p^0)^2-\vp^2)= {\mathcal I}_2
((p^0)^2-\vp^2)$, so that we obtain the $\theta(p^0)$ factor
in Eq. (\ref{BB14a}) via $\theta(p^0) = {1\over 2} (1+ \epsilon(p^0))$.
The result ${\mathcal I}_1 ((p^0)^2-\vp^2)= {\mathcal I}_2
((p^0)^2-\vp^2)$ carries over to the finite temperature case
with the shift $\vp \rightarrow \vq = \vp +\vp_b$, so that
we may also write ${\mathcal I} = {\mathcal I}_1 + {\mathcal I}_2
= \theta (p^0) \, 2 {\mathcal I}_2$.

We now use this result back in Eq. (\ref{BB9}) to carry out the integration over
$s'$, leaving the integration over $\vp_b$:
%
\beqar
W(x, y) &=& {1\over \Z} \int \frac{d^4 p}{(2 \pi)^4}\,e^{- i p\cdot (x-y)} 
\int
{d^3 p_b}\label{BB16}\\
&&\times 
\int_0^\infty {ds'' \over 2 p^0_b (2\pi)^3} e^{-\beta p^0_b} \, {\mathcal I} ( (p^0)^2 - \vq^2,
s'') \; .
\nonumber
\eeqar
At this point, we see that, apart
from $\vq = \vp + \vp_b$, the integrand can depend on
$\vp_b^2$ via $p^0_b$ as well, due to the exponential
factor and the integration measure in Eq. (\ref{BB16}).
So we define a spectral function
\beq
\rho (\vp_b^2, (p^0)^2 - \vq^2 ) = {1\over \Z} 
\int_0^\infty {ds'' \over 2 p^0_b (2\pi)^3}
e^{- \beta p^0_b} \, 2\,{\mathcal I}_2 ((p^0)^2 - \vq^2 , s'') \; .
\label{BB17}
\eeq
We can then write $W(x,y)$ as
\beq
W(x, y) = \int \frac{d^4 p}{(2 \pi)^4} e^{- i p\cdot (x-y)} \theta(p^0)
\int
{d^3 p_b}\,
\rho (\vp_b^2, (p^0)^2 - \vq^2 ) \; .
\label{BB18}
\eeq

Consider now the integration over $p^0$. For positive values of
$p^0$, we define a variable $s$ by $p^0 = \sqrt{s +\vq^2}$ and change variables to $s$. For a function $f(p^0)$ which is even
in $p^0$, the integral can then be written as
\beqar
&&\int_{-\infty}^\infty dp^0 \, e^{ i p^0 (x^0 - y^0) } \, f(p^0)
\label{BB19}
\\
&&= \int_0^\infty dp^0 \, \left[e^{ i p^0 (x^0 - y^0) }
+ e^{ - i p^0 (x^0 - y^0) } \right] \, f(p^0)\nonumber\\
&&= \int_0^\infty {ds \over  2\sqrt{s+ \vq^2}} 
 \left[e^{ i p^0 (x^0 - y^0) }
+ e^{ - i p^0 (x^0 - y^0) } \right] \, f(p^0)
\nonumber\\
&&= \int_0^\infty ds \int_{-\infty}^\infty d p^0\delta ( (p^0)^2 - \vq^2 - s) 
e^{i p^0 (x^0 - y^0) } f(p^0) \; . \nonumber
\eeqar
The second line where we consider the range of integration to be
in the range $[0,\infty]$ is necessary to use $s$ in place of
$p^0$.
In the third line of this equation, it is understood that $p^0$ in the exponent
and the argument of $f(p^0)$ is $\sqrt{s + \vq^2}$. 
In the last line, we introduce another 
$p^0$ as a free variable of integration.
In a similar way
\beqar
&&\int_{-\infty}^\infty dp^0 \, e^{ i p^0 (x^0 - y^0) } \, \epsilon(p^0)\,f(p^0)
\label{BB20} \\
&&= \int_0^\infty ds \int_{-\infty}^\infty d p^0 \epsilon (p^0)\delta ( (p^0)^2 - \vq^2 - s) 
e^{i p^0 (x^0 - y^0) } f(p^0) \; . \nonumber
\eeqar
These equations can be combined to make a similar statement about
an even function multiplied by $\theta (p^0)$.
Going back to (\ref{BB18}) and using these results, we get
\beqar
W(x, y)&=& \int \frac{d^4 p}{(2 \pi)^4}\,e^{- i p\cdot (x-y)} 
\int {d^3 p_b} \int_0^\infty ds
\label{BB21}\\
&&\hskip .2in \theta(p^0)\delta \bigl(s - (p^0)^2 + (\vp +\vp_b)^2 \bigr)
\rho ( \vp_b^2, s ) \; . \nonumber
\eeqar
The contribution from ${\mathcal I}_1$ is symmetric in
$(x^0 - y^0)$ and does not contribute.
The term involving ${\mathcal I}_2$ does contribute;
it has a factor of $\epsilon (p^0)$ in the integrand as well.
This is basically the spectral representation
derived by Bros and Buchholz
\cite{Bros:1992ey,Bros:1994ofl,Bros:1996mw,Bros:2001zs}.

A more explicit form is obtained if we use the KMS condition for
the Wightman function. In terms of its Fourier transform ${\widetilde W}(p)$, the
KMS condition implies that
\begin{equation}
    {\widetilde W} (p) = e^{\beta p^0} \, {\widetilde W}(-p)
    = {{\widetilde W}(p) - {\widetilde W}(-p)\over 1- e^{-\beta p^0}}
    \label{BB22}
\end{equation}
From Eq. (\ref{BB21}), we see that
\beqar
    {\widetilde W}(p) - {\widetilde W}(-p) &=& \int d^3p_b\; \int_0^\infty ds \;\epsilon(p^0)\nonumber\\
    &&
    \delta \bigl(s - (p^0)^2 + (\vp +\vp_b)^2 \bigr)
    \rho(\vp_b^2, s) ,
    \label{BB23}
\eeqar
so that we can write the spectral representation for the Fourier transform of $W(x, y)$ as 
\begin{align}
    {\widetilde W}(p) =&
    \nonumber \\  
    {\epsilon (p^0)\over 1- e^{ -\beta p^0}}&
    \int_0^\infty ds\, \int d^3p_b \, \delta \bigl(s - (p^0)^2 + (\vp -\vp_b)^2 \bigr)
\rho ( \vp_b^2, s )
\label{BB24}
\end{align}
In this equation, we have changed $\vp_b $ to $- \vp_b$ to facilitate comparison with \cite{Bros:1992ey,Bros:1994ofl,Bros:1996mw,Bros:2001zs}.
Notice that, as $\beta \rightarrow \infty$, $(1- e^{-\beta p^0})^{-1}
\rightarrow \theta(p^0)$ and
$\rho(\vp_b^2, s) \rightarrow \rho(0,s)$ because of the 
$e^{-\beta p_b^0}$-factor. Thus from Eq. (\ref{BB24}) we 
correctly recover the vacuum spectral representation
in this limit.

While we have used Lorentz invariance of the underlying dynamics to 
bring the spectral representation to the form given above,
the final result obviously is not Lorentz-invariant, since the
density matrix represents the choice of state in the rest frame of
the medium. This is evident from the
term $e^{-\beta p_b^0}$ and the partition function $\Z$ in the expression for
the spectral function.

It is straightforward to generalize the result to a moving medium.
We replace 
\begin{equation}
{\rm e}^{- \beta H} \rightarrow {\rm e}^{ - \beta (H u^0 - \vu\cdot {\vec P})}
\; , 
\end{equation}
where 
\begin{equation}
u^\mu = \frac{1}{\sqrt{1- v^2}} (1, v^i) \; ,
\end{equation}
is the 4-velocity of the medium. Similarly,
\begin{equation}
{\rm e}^{- \beta p_b^0} \rightarrow {\rm e}^{-\beta (p_b^0 u^0 - \vp_b\cdot \vu )}
\; , 
\end{equation} 
and likewise for the partition function.
With these changes, (\ref{BB21})
gives the spectral representation of 
$W(x,y)$ in a moving medium.

\section{Conclusions}

In this paper we have given a (relatively) simple derivation of the
Bros-Buchholz form of the spectral density for a scalar field at
nonzero temperature.  While relativistic invariance in vacuum
constrains this form, it is still a function fo two variables.

At nonzero temperature, the difficulty in measuring the spectral density
is that one only has the values at discrete points, which are much
sparser than the values for spatial momenta.  We hope that our derivation
will inspire others to find a greater utility in this form.

\acknowledgments
We thank P. Lowdon for discussions and suggestions, especially 
for pointing out the simplification using the KMS condition.
R.D.P. also thanks O. Philipsen for numerous discussions.
V.P.N. was supported in part by the U.S. National Science Foundation Grants No. PHY-2112729 and PHY-2412479.
R.D.P. was supported by the U.S. Department of Energy under contract DE-SC0012704 and thanks the Alexander v. Humboldt Foundation 
for their support.

%

\end{document}